\documentclass[preprint,aps ,nofootinbib]{revtex4}
\pdfoutput=1
\usepackage{graphicx}%Include figure files
\usepackage{amsmath}
\usepackage{amsfonts}
\usepackage{amssymb}
\usepackage{color}%
\usepackage{dcolumn}% Align table columns on decimal point
\providecommand{\U}[1]{\protect\rule{.1in}{.1in}}

\newcommand{\f}{\begin{equation}}
\newcommand{\ff}{\end{equation}}
\newcommand{\fa}{\begin{eqnarray}}
\newcommand{\ffa}{\end{eqnarray}}

\begin{document}
\title{Holographic fermions in charged Gauss-Bonnet black hole}
\author{Jian-Pin Wu}
\email{jianpinwu@yahoo.com.cn}
\affiliation{Department of Physics, Beijing Normal University, Beijing 100875, China}

\begin{abstract}

We study the properties of the Green's functions of the fermions in charged Gauss-Bonnet black hole.
What we want to do is to investigate how the presence of Gauss-Bonnet coupling constant $\alpha$ affects
the dispersion relation, which is a characteristic of Fermi or non-Fermi liquid,
as well as what properties such a system has, for instance, the Particle-hole (a)symmetry.
One important result of this research is that we find for $q=1$,
the behavior of this system is different from that of the Landau Fermi liquid
and so the system can be candidates for holographic dual of generalized non-Fermi liquids.
More importantly, the behavior of this system increasingly similar to that of the Landau Fermi liquid
when $\alpha$ is approaching its lower bound.
Also we find that this system possesses the Particle-hole asymmetry when $q\neq 0$,
another important characteristic of this system.
In addition, we also investigate briefly the cases of the charge dependence.

\end{abstract} \maketitle

\section {Introduction}

As is known to all, the success of the single electron picture of metals rests on Landau's Fermi-liquid theory.
The metal in such theory is treated as a gas of Fermi particles whose interactions are weak and
not as significant as that of the original electrons. The reason is that particles of this model are not
the original electrons but the electron-like quasi-particles that emerge from the interacting gas of electrons.
However, recently some new materials, including the cuprate superconductors
and other oxides, seem to lie outside this framework.
For these new materials, we refer them to the non-Fermi liquid metals.
They show lots of new physical properties which can not
be understood in terms of weakly interacting electron-like objects.
For the non-Fermi liquid, a sharp Fermi surface still exists.
But quasi-particle picture breaks down generically.
Although there have been many phenomenological models to describe the non-Fermi liquid,
a general theoretical framework characterizing non-Fermi liquid metals remains a suspense.
Therefore, it is necessary to develop a basic principle for non-Fermi liquid.
Maybe the AdS/CFT correspondence can provide us with a possible clue to yield the basic principle of non-Fermi liquid.
Indeed, by applying AdS/CFT correspondence \cite{Maldacena1997,Gubser1998,Witten1998}, some breakthroughs have been achieved, that is, we find some new classes of non-Fermi liquids \cite{HongLiuNon-Fermi}.
Varieties of holographic fermions models and their extensions are being explored.
These can be seen from Refs.
\cite{StringQEFL,FermionsBTZBH,HNFMagneticFieldBasu,StrangeMetallicHartnoll,SemiHFLPolchinski,HNFLMagneticFGubankova,HFLDynamicalGap,HFLDipoleCoupling}.

However, before string theory is fully understood,
it is necessary to consider the higher curvature (or derivative) interactions in an effective gravitational theory.
From the point of view of the AdS/CFT, the higher curvature interactions on the gravity side correspond to finite
coupling corrections on the gauge theory side,
thus broadening the class of field theories one can holographically study.
The main motivation of considering
these corrections derives from the fact that string theory contains higher
curvature corrections arising from stringy effects.
A simple and useful model with regard these corrections is Gauss-Bonnet (GB) gravity,
which contains only the curvature-squared interaction.
Several models of holographic superconductor have been exploited in this setting \cite{GBHS1,GBHS2,GBHS3,GBHS4,GBHS5,GBHS6,GBHS7}.
Recently, a new higher derivative theory of gravity (quasi-topological gravity) is constructed,
which contains not only the curvature-squared interaction but also a curvature-cubed interaction \cite{QT,QTblackhole}.
The corresponding holographic superconductor models have also been discussed in \cite{HSinQT1,HSinQT2}.
Another interesting extension comes from the coupling between the Maxwell field and the bulk Weyl tensor \cite{Weyl1,Weyl2}.
Also, in Refs. \cite{Weyl2,WeylHS}, they find that the Weyl corrections can describe the dual field theories with both the weak-coupling and the strong-coupling, which is a very interesting and significant holographic study on the dual field theories.
Therefore, it is deserving to further exploit the effect of higher curvature interactions on fermions by using AdS/CFT correspondence step by step.
In this paper, we consider the Gauss-Bonnet term as the first step to introduce the stringy correction into the gravitational action.
And we will mainly focus on how the GB coupling constant affects the spectral function of the fermions.

Our paper is organized as follows.
In section II, we briefly introduce the charged Gauss-Bonnet black hole.
Following Ref. \cite{HongLiuNon-Fermi}, we obtain the Dirac equation of the probe fermions in the charged Gauss-Bonnet black hole in section III.
The numerical results are presented in section IV
where we will mainly focus on the dispersion and the Particle-hole asymmetry in particular.
Conclusions and discussions follow in section VI.
Finally, in appendix A, we give a brief summary on the analytic treatment
in the low frequency limit developed in Ref.\cite{HongLiuAdS2}.

\section {Charged Black Holes in Gauss-Bonnet gravity}\label{CBHinGBG}

In this section, a brief review of black holes in Gauss-Bonnet gravity is given as follows.
Conventionally, we begin with the following action coupled a vector field $A_{a}$
\footnote{Here we have set the the five dimensional gravitational constant $\kappa_{5}^{2}=1/2$
and the effective dimensionless gauge coupling $g_{F}=2$.}:
\begin{eqnarray}
\label{GBaction}
S=\int d^{5}x \sqrt{-g}\left[R+\frac{12}{L^{2}}+\frac{\alpha}{2}\left(R^{abcd}R_{abcd}-4R^{ab}R_{ab}+R^{2}\right)-\frac{1}{4}F^{ab}F_{ab}\right],
\end{eqnarray}
where $R_{abcd}$, $R_{ab}$ and $R$ are the Riemann curvature tensor, Ricci tensor,
and the Ricci scalar, respectively.
$L$ is the AdS radius and for convenience, we will set $L=1$.
$\alpha$ is Gauss-Bonnet coupling constant and the constraints on it will also be discussed in the following parts of this paper.
And as is commonly known, $F_{ab}=\partial_{a}A_{b}-\partial_{b}A_{a}$.
By applying the principle of variation to (\ref{GBaction}), we can easily obtain the equations of motion:

Maxwell¡¯s equations
\begin{eqnarray}
\label{MaxwellE}
\nabla_{a}F^{ab}=0,
\end{eqnarray}
and Einstein¡¯s equations
\begin{eqnarray}
\label{EinsteinE}
R_{ab}-\frac{1}{2}Rg_{ab}+6g_{ab}-\alpha\left[H_{ab}-\frac{1}{4}Hg_{ab}\right]
=\frac{1}{2}F_{ac}F_{b}^{c}-\frac{1}{8}F_{cd}^{2}g_{ab},
\end{eqnarray}
where
\begin{eqnarray}
\label{TensorHab}
H_{ab}=R_{a}^{cde}R_{bcde}-2R_{ac}R_{b}^{c}-2R_{acbd}R^{cd}+RR_{ab}.
\end{eqnarray}
and
\begin{eqnarray}
\label{TensorHtrace}
H=H_{a}^{a}.
\end{eqnarray}

Under a certain circumstance in which we take the following metric with planar symmetry
\begin{eqnarray}
\label{MetricA}
ds^{2}=-f(r)dt^{2}+\frac{dr^{2}}{f(r)}+r^{2}(dx^{2}+dy^{2}+dz^{2}),
\end{eqnarray}
and the ansatz of the gauge fields as
\begin{eqnarray}
\label{GaugeFA}
A_{a}=(A_{t}(r),0,0,0),
\end{eqnarray}
the equations of motion (\ref{MaxwellE}) and (\ref{EinsteinE}) reduce to
\begin{eqnarray}
\label{MaxwellESimplify}
A''_{t}+\frac{3}{r}A'_{t}=0,
\end{eqnarray}
\begin{eqnarray}
\label{EinsteinESimplify}
\left(1-\frac{2\alpha f}{r^{2}}\right)f'+\frac{2}{r}f-4r
+\frac{r A'^{2}_{t}}{6}=0,
\end{eqnarray}
where the prime represents derivative with respect to $r$.
The solutions of the above equations are  \cite{GBblackhole1,GBblackhole2}
\begin{eqnarray}
\label{metricf}
f(r)=\frac{r^{2}}{2\alpha}\left[1-\sqrt{1-4\alpha\left(1-\frac{r_{+}^{4}}{r^{4}}\right)+\frac{4\alpha \mu^{2}r_{+}^{2}}{3r^{4}}\left(1-\frac{r_{+}^{2}}{r^{2}}\right)}\right],
\end{eqnarray}
and
\begin{eqnarray}
\label{metricA}
A_{t}=\mu\left(1-\frac{r_{+}^{2}}{r^{2}}\right),
\end{eqnarray}
where $r_{+}$ is the horizon radius determined by $f(r_{+})=0$\footnote{In the following, for convenience, we will set $r_{+}=1$.}, and $\mu$ can be identified with the chemical potential of the dual field theory.
In the Einstein limit $\alpha\rightarrow 0$ the formula above reduces to the Reissner-Nordstr$\ddot{o}$m
AdS black hole
\begin{eqnarray}
\label{metricfRN}
f(r)|_{\alpha\rightarrow 0}=r^{2}\left[\left(1-\frac{r_{+}^{4}}{r^{4}}\right)-\frac{\mu^{2}r_{+}^{2}}{3r^{4}}\left(1-\frac{r_{+}^{2}}{r^{2}}\right)\right].
\end{eqnarray}

By using the standard approach of euclidean continuation
near the black hole horizon, the Hawking temperature of the black hole in Gauss-Bonnet gravity is
\begin{eqnarray}
\label{HawkingT}
T=\frac{f'(r_{+})}{4\pi}=\frac{1}{\pi}\left(1-\frac{\mu^{2}}{6}\right),
\end{eqnarray}
which is also the temperature of the conformal field theory on the boundary of the AdS
spacetime. It is independent of the Gauss-Bonnet coupling constant $\alpha$.

In the following, we will discuss the behaviors of places verging on the boundary ($r\rightarrow \infty$) and the horizon ($r\rightarrow 1$), respectively.
Near the boundary of the bulk, the redshift factor $f(r)$ becomes
\begin{eqnarray}
\label{metricfinfty}
f_{\infty}\equiv \lim_{r\rightarrow \infty}f(r)=\frac{r^{2}}{2\alpha}\left[1-\sqrt{1-4\alpha}\right]=r^{2}C_{\alpha},
\end{eqnarray}
where we denote $C_{\alpha}\equiv(1-\sqrt{1-4\alpha})/2\alpha$.
Then, the geometry of the black hole on the boundary can be reexpressed as
\begin{eqnarray}
\label{MetricB}
ds^{2}=-r^{2}C_{\alpha}dt^{2}+\frac{dr^{2}}{r^{2}C_{\alpha}}+r^{2}(dx^{2}+dy^{2}+dz^{2}).
\end{eqnarray}

It is also pure $AdS_{5}$, but depends on the GB coupling constant $\alpha$.
In order to have a well-defined anti-de Sitter vacuum for the gravity theory,
a certain condition $\alpha\leq 1/4$ should be required,
whose upper bound $\alpha=1/4$ is known as the Chern-Simons limit.
Furthermore, considering the causality of dual field theory on the boundary,
there exists a stronger constraint on the GB coupling in five dimensions
\cite{GBcouplingConstraint1,GBcouplingConstraint2,GBcouplingConstraint3,GBcouplingConstraint4,GBcouplingConstraint5,
GBcouplingConstraint6,GBcouplingConstraint7,GBcouplingConstraint8,GBcouplingConstraint9,GBcouplingConstraint10}
\begin{eqnarray}
\label{GBcouplingConstraint}
-\frac{7}{36}\leq \alpha\leq \frac{9}{100}.
\end{eqnarray}

As for the condition of the horizon, we in reality do a similar job, but for this part, we consider a special case, that is, the zero-temperature
which means $\mu=\sqrt{6}$.
Then, the redshift factor $f(r)$ becomes
\begin{eqnarray}
\label{metricfT0}
f(r)|_{T=0}=\frac{r^{2}}{2\alpha}\left[1-\sqrt{1-4\alpha+\frac{12\alpha}{r^{4}}-\frac{8\alpha}{r^{6}}}\right].
\end{eqnarray}

Obviously, when $r\rightarrow 1$, $f(r)\approx 12(r-1)^{2}$, which is also independent of the Gauss-Bonnet coupling constant $\alpha$.
In light of this, near the horizon, the geometry is no longer the pure $AdS_{5}$, but the $AdS_{2}\times \mathbb{R}^{3}$ with the curvature radius of
$AdS_{2}$, $L_{2}=\frac{1}{\sqrt{12}}L$.

\section {Dirac equation}

Now, we consider probe fermions in the charged Gauss-Bonnet black hole.
We have the bulk fermion action as following \cite{HongLiuNon-Fermi}
\begin{eqnarray}
\label{actionspinor}
S_{D}=i\int d^{5}x \sqrt{-g}\overline{\zeta}\left(\Gamma^{a}\mathcal{D}_{a}-m_{\zeta}\right)\zeta,
\end{eqnarray}
where $\mathcal{D}_{a}$ is the covariant derivative given by\footnote{Throughout the paper,
we will use the conventions of \cite{Conventions1}. $a$, $b$ are the usual spacetime abstract index
and $\mu$, $\nu$ are the tangent-space index.}$^{,}$\footnote{Note that the Dirac action (\ref{actionspinor}),
only through the effective chemical potential $\mu_{q}\equiv \mu q$, that is to say,
through the combination of $g_{F}q$, depends on $q$. For convenience, the $g_{F}$ has been set as $2$ above
and the $q$ is treated as a free parameter,
but we should note that only the product of them is the relevant quantity.
For more discussions, please refer to Ref.\cite{HongLiuNon-Fermi}.
At the same time, we also want to remind readers to pay attention to the fact that
in Ref.\cite{HongLiuNon-Fermi,HongLiuAdS2}, they set $g_{F}=1$, which is different from our conventions.}
\begin{eqnarray}
\label{Dderivative}
\mathcal{D}_{a}=\partial_{a}+\frac{1}{4}(\omega_{\mu\nu})_{a}\Gamma^{\mu\nu}-iqA_{a},
\end{eqnarray}
$(\omega_{\mu\nu})_{a}$ is the spin connection 1-forms given by
\begin{eqnarray}
\label{spinconnectionD}
(\omega_{\mu\nu})_{a}=(e_{\mu})^{b}\nabla_{a}(e_{\nu})_{b},
\end{eqnarray}
and
\begin{eqnarray}
\label{spinconnection}
\Gamma^{\mu\nu}=\frac{1}{2}[\Gamma^{\mu},\Gamma^{\nu}],~~~~~~\Gamma^{a}=(e_{\mu})^{a}\Gamma^{\mu},
\end{eqnarray}
where $(e_{\mu})^{a}$ form a set of orthogonal normal
vector bases. The Dirac equation derived from the action $S_{D}$ is expressed as
\begin{eqnarray}
\label{DiracEquation1}
\Gamma^{a}\mathcal{D}_{a}\zeta-m_{\zeta}\zeta=0.
\end{eqnarray}

We should choose the following orthogonal normal vector bases
\begin{eqnarray}
\label{VectorBases}
(e_{\mu})^{a}=\sqrt{g^{\mu\mu}}(\frac{\partial}{\partial
\mu})^{a},~~\mu=t,x,y,z,r.
\end{eqnarray}

Using (\ref{spinconnectionD}), one can calculate the non-vanishing
components of spin connections as follows
\begin{eqnarray}
\label{SpinConnections}
(\omega_{tr})_{a} &=& -(\omega_{rt})_{a}=-\sqrt{g^{rr}}\partial_{r}(\sqrt{g_{tt}})(dt)_{a},
\nonumber\\ \label{S1}(\omega_{ir})_{a} &=& -(\omega_{ri})_{a}=-\sqrt{g^{rr}}\partial_{r}(\sqrt{g_{ii}})(dx^{i})_{a},~~i=x,y,z.
\end{eqnarray}

Following Ref.\cite{HongLiuNon-Fermi}, we can make a transformation
$\zeta=(-g g^{rr})^{-\frac{1}{4}}\mathcal{F}$ to remove the spin
connection in Dirac equation. Then, the equation turns out to be
\begin{eqnarray}
\label{DiracEquation2}
\sqrt{g^{rr}}\Gamma^{r}\partial_{r}\mathcal{F}
+\sqrt{g^{tt}}\Gamma^{t}(\partial_{t}-iq A_{t})\mathcal{F}
+\left(\sum_{i}\sqrt{g^{ii}}\Gamma^{i}\partial_{i}\right)\mathcal{F}
-m_{\zeta}\mathcal{F}=0.
\end{eqnarray}

Next, we will work in Fourier space where we expand $\mathcal{F}$ as $\mathcal{F}=F e^{-i\omega t +ik_{i}x^{i}}$.
Then, the Dirac equation can be rewritten as
\begin{eqnarray}
\label{DiracEinFourier}
\sqrt{g^{rr}}\Gamma^{r}\partial_{r}F
-i(\omega+q)\sqrt{g^{tt}}\Gamma^{t}F
+i k \sqrt{g^{xx}}\Gamma^{x}F
-m_{\zeta}F=0.
\end{eqnarray}
where due to rotational symmetry in $x-y-z$ directions, we set $k_{y}=k_{z}=0$ and $k_{x}=k$ without losing generality.
We will choose the following basis for our gamma matrices as in\cite{HongLiuAdS2,Photoemission}
\begin{eqnarray}
\label{GammaMatrices}
 && \Gamma^{r} = \left( \begin{array}{cc}
-\sigma^3  & 0  \\
0 & -\sigma^3
\end{array} \right), \;\;
 \Gamma^{t} = \left( \begin{array}{cc}
 i \sigma^1  & 0  \\
0 & i \sigma^1
\end{array} \right),  \;\;
\Gamma^{x} = \left( \begin{array}{cc}
-\sigma^2  & 0  \\
0 & \sigma^2
\end{array} \right),
\qquad \ldots
\end{eqnarray}

Splitting the 4-component spinors into two 2-component spinors $F=(F_{1},F_{2})^{T}$,
we have a new version of the Dirac equation as
\begin{eqnarray} \label{DiracEF}
\sqrt{g^{rr}}\partial_{r}\left( \begin{matrix} F_{1} \cr  F_{2} \end{matrix}\right)
+m_{\zeta}\sigma^3\otimes\left( \begin{matrix} F_{1} \cr  F_{2} \end{matrix}\right)
=\sqrt{g^{tt}}(\omega+qA_{t})i\sigma^2\otimes\left( \begin{matrix} F_{1} \cr  F_{2} \end{matrix}\right)
\mp  k \sqrt{g^{xx}}\sigma^1 \otimes \left( \begin{matrix} F_{1} \cr  F_{2} \end{matrix}\right)
~,
\end{eqnarray}
a decouple equation between $F_{1}$ and $F_{2}$.
After achieving this equation, we will discuss it in a special case, that is, near the boundary ($r\rightarrow \infty$) in order to acquire some information about the Green function near the boundary. From the forgoing information, we know that in this special case, the geometry is an an asymptotic $AdS_{5}$, $g_{rr}\approx\frac{1}{r^{2}}$ and $g_{ii}\approx r^{2},~~~i=t,x,y,z$.
Under such a condition, Eq. (\ref{DiracEF}) becomes
\begin{eqnarray} \label{DiracEboundary}
(r\partial_{r}+m_{\zeta}\sigma^3)\otimes \left( \begin{matrix} F_{1} \cr  F_{2} \end{matrix}\right)=0~,
\end{eqnarray}
another decouple equation between $F_{1}$ and $F_{2}$ whose solution can be expressed as
\begin{eqnarray} \label{BoundaryBehaviour}
F_{\alpha} \buildrel{r \to \infty}\over {\approx} a_{\alpha}r^{m_{\zeta}}\left( \begin{matrix} 0 \cr  1 \end{matrix}\right)
+b_{\alpha}r^{-m_{\zeta}}\left( \begin{matrix} 1 \cr  0 \end{matrix}\right),
\qquad
\alpha = 1,2~.
\end{eqnarray}

If $b_{\alpha}\left( \begin{matrix} 1 \cr  0 \end{matrix}\right)$ and $a_{\alpha}\left( \begin{matrix} 0 \cr  1 \end{matrix}\right)$
are related by
\begin{eqnarray} \label{EVEsourceRelation}
b_{\alpha}\left( \begin{matrix} 1 \cr  0 \end{matrix}\right)=\mathcal{S}a_{\alpha}\left( \begin{matrix} 0 \cr  1 \end{matrix}\right)~,
\end{eqnarray}
then the boundary spinor Green functions $G$ is given by \cite{HongLiuSpinor}
\begin{eqnarray} \label{Grgamma}
G=-i \mathcal{S}\gamma^{0}~.
\end{eqnarray}
where $\gamma^{0}$ is the gamma matrices of the boundary theory and $\gamma^{0}=i\sigma^1$.
In order to find the matrix $\mathcal{S}$, we will make such a decomposition $F_{\pm}=\frac{1}{2}(1\pm \Gamma^{r})F$
according to eigenvalues of $\Gamma^{r}$. Then
\begin{eqnarray} \label{gammarDecompose}
F_{+}=\left( \begin{matrix} \mathcal{B}_{1} \cr  \mathcal{B}_{2} \end{matrix}\right),~~~~~~~F_{-}=\left( \begin{matrix} \mathcal{A}_{1} \cr  \mathcal{A}_{2} \end{matrix}\right),
\end{eqnarray}
with
\begin{eqnarray} \label{gammarDecompose}
F_{\alpha} \equiv \left( \begin{matrix} \mathcal{A}_{\alpha} \cr  \mathcal{B}_{\alpha} \end{matrix}\right).
\end{eqnarray}

Now, by using (\ref{BoundaryBehaviour}), (\ref{EVEsourceRelation}) and (\ref{Grgamma}), we can express the boundary Green functions as following
\begin{eqnarray} \label{GreenFBoundary}
G (\omega,k)= \lim_{r\rightarrow \infty} r^{2m_{\zeta}} \widetilde{G}(r,\omega,k),
\end{eqnarray}
here we have defined the following matrices:
\begin{eqnarray} \label{AB}
\widetilde{G}(r,\omega,k)\equiv \left( \begin{array}{cc}
\widetilde{G}_{11}   & 0  \\
0  & \widetilde{G}_{22} \end{array} \right)  \ ,
\end{eqnarray}
and
\begin{eqnarray} \label{GAB}
\widetilde{G}_{\alpha\alpha}(r,\omega,k)\equiv \frac{\mathcal{A}_{\alpha}}{\mathcal{B}_{\alpha}}~~~,\alpha=1,2.
\end{eqnarray}

Moreover, under the same decomposition, the Dirac equation (\ref{DiracEF}) can be rewritten as
\begin{eqnarray} \label{DiracEAB1}
(\sqrt{g^{rr}}\partial_{r}\pm m_{\zeta})\left( \begin{matrix} \mathcal{A}_{1} \cr  \mathcal{B}_{1} \end{matrix}\right)
=\pm(\omega+qA_{t})\sqrt{g^{tt}}\left( \begin{matrix} \mathcal{B}_{1} \cr  \mathcal{A}_{1} \end{matrix}\right)
-k \sqrt{g^{xx}} \left( \begin{matrix} \mathcal{B}_{1} \cr  \mathcal{A}_{1} \end{matrix}\right)
~,
\end{eqnarray}
\begin{eqnarray} \label{DiracEAB2}
(\sqrt{g^{rr}}\partial_{r}\pm m_{\zeta})\left( \begin{matrix} \mathcal{A}_{2} \cr  \mathcal{B}_{2} \end{matrix}\right)
=\pm(\omega+qA_{t})\sqrt{g^{tt}}\left( \begin{matrix} \mathcal{B}_{2} \cr  \mathcal{A}_{2} \end{matrix}\right)
+k \sqrt{g^{xx}} \left( \begin{matrix} \mathcal{B}_{2} \cr  \mathcal{A}_{2} \end{matrix}\right)
~.
\end{eqnarray}

Using the method developed in \cite{HongLiuUniversality,HongLiuSpinor,HongLiuAdS2},
one can package the Dirac equation (\ref{DiracEAB1}) and (\ref{DiracEAB2}) into the evolution equation of $\widetilde{G}(r,\omega,k)$,
which will be more convenient to impose the initial conditions at the horizon and read off the boundary Green functions,
\begin{eqnarray} \label{DiracEF1}
(\sqrt{g^{rr}}\partial_{r}
+2m_{\zeta})\widetilde{G}
=\widetilde{G}\left(\sqrt{g^{tt}}(\omega+qA_{t})+k \sqrt{g^{xx}}\sigma^{3}\right)\widetilde{G}
+(\sqrt{g^{tt}}(\omega+qA_{t})-k \sqrt{g^{xx}}\sigma^{3})~.
\end{eqnarray}

The boundary condition of the matrix $\widetilde{G}(r,\omega,k)$ in this new equation
\begin{eqnarray} \label{GatTip}
\widetilde{G}_{\alpha\alpha}(r,\omega,k)\buildrel{r \to 1}\over =i.
\end{eqnarray}
can be derived from the
requirement that the solutions of Eqs. (\ref{DiracEAB1}) and (\ref{DiracEAB2}) at the horizon, $r\rightarrow 1$,
be in-falling.

However, we also note that for $T=0$ and $\omega=0$, the boundary condition has to be modified as follows
\begin{eqnarray} \label{GhorizonTw0}
\widetilde{G}_{\alpha\alpha}(r,\omega=0,k)\buildrel{r \to 1}\over =\frac{m_{\zeta}-\sqrt{m_{\zeta}^{2}+k^{2}-\frac{\mu_{q}^{2}}{12}-i\epsilon}}{k+\frac{\mu_{q}}{\sqrt{12}}}.
\end{eqnarray}

\section{Properties of spectral functions}

\subsection{General behavior} \label{GeneralB}

As Ref.\cite{HongLiuNon-Fermi}, we also have some symmetry properties of the Green function by direct inspection of the equation (\ref{DiracEF1}).
We itemize them as follows:

(1) $G_{22}(\omega,k)=G_{11}(\omega,-k)$;~~~~(2) $G_{22}(\omega,k;-q)=G_{11}^{\ast}(-\omega,k;q)$;~~~~

For the case $m_{\zeta}=0$,

(3) $G_{22}(\omega,k)=-\frac{1}{G_{11}(\omega,k)}$;~~~~(4) $G_{22}(\omega,k=0)=G_{11}(\omega,k=0)=i$.

When the background geometry is pure $AdS_{5}$ (Eq. (\ref{MetricB})),
for massless bulk fermion, the Dirac equation (\ref{DiracEF1}) can be explicitly expressed as
\begin{eqnarray} \label{DiracEFpureAdS1}
r^{2}\partial_{r}\mathcal{G}_{11}=\left(\frac{\omega+\mu_{q}}{C_{\alpha}}+\frac{k}{C_{\alpha}^{1/2}}\right)\mathcal{G}_{11}^{2}
+\frac{\omega+\mu_{q}}{C_{\alpha}}-\frac{k}{C_{\alpha}^{1/2}},
\end{eqnarray}
\begin{eqnarray} \label{DiracEFpureAdS2}
r^{2}\partial_{r}\mathcal{G}_{22}=\left(\frac{\omega+\mu_{q}}{C_{\alpha}}-\frac{k}{C_{\alpha}^{1/2}}\right)\mathcal{G}_{22}^{2}
+\frac{\omega+\mu_{q}}{C_{\alpha}}+\frac{k}{C_{\alpha}^{1/2}},
\end{eqnarray}
where we denote the Green function in pure $AdS_{5}$ background as $\mathcal{G}$.
The solution of the above equations can be easily obtained as
\footnote{We can also refer to Refs.\cite{GreenFpureAdS1,GreenFpureAdS2}.}
\begin{eqnarray} \label{GreenFpureAdS}
\mathcal{G}_{11}=-\sqrt{\frac{(\omega+\mu_{q})/C_{\alpha}-k/C_{\alpha}^{1/2}+i\epsilon}{(\omega+\mu_{q})/C_{\alpha}+k/C_{\alpha}^{1/2}+i\epsilon}},~~~
\mathcal{G}_{22}=\sqrt{\frac{(\omega+\mu_{q})/C_{\alpha}+k/C_{\alpha}^{1/2}+i\epsilon}{(\omega+\mu_{q})/C_{\alpha}-k/C_{\alpha}^{1/2}+i\epsilon}},
\end{eqnarray}
where $\epsilon\rightarrow 0$.
For $q=0$, it is clear that the spectral function has a Particle-hole symmetry (symmetry under $(\omega,k)\rightarrow (-\omega,-k)$).
However, when $q\neq 0$, the Particle-hole symmetry is broken (Particle-hole asymmetry).
In addition, the spectral function $Im \mathcal{G}$ has also an edge-singularity along $\omega=\pm k$
and vanishs in the region $\omega\in (-k,k)$.

Now, we turn to the charged Gauss-Bonnet black hole background (Eqs. (\ref{MetricA}) and (\ref{metricf})).
In this case, the Dirac equation (\ref{DiracEF1}) becomes
\begin{eqnarray} \label{DiracEFCGB1}
f^{1/2}\partial_{r}\widetilde{G}_{11}+2m_{\zeta}\widetilde{G}_{11}
=\left[\frac{1}{f^{1/2}}(\omega+qA_{t})+k\frac{1}{r}\right]\widetilde{G}_{11}^{2}
+\frac{1}{f^{1/2}}(\omega+qA_{t})-k\frac{1}{r},
\end{eqnarray}
\begin{eqnarray} \label{DiracEFCGB2}
f^{1/2}\partial_{r}\widetilde{G}_{22}+2m_{\zeta}\widetilde{G}_{22}
=\left[\frac{1}{f^{1/2}}(\omega+qA_{t})-k\frac{1}{r}\right]\widetilde{G}_{22}^{2}
+\frac{1}{f^{1/2}}(\omega+qA_{t})+k\frac{1}{r}.
\end{eqnarray}

We can solve the above equations numerically with
the boundary conditions (\ref{GatTip}) to investigate the properties of the spectral function.
In this paper, we will only focus on the massless fermion ($m_{\zeta}=0$) and extremal charged GB black hole (zero temperature limit).
The dependence of the spectral function on mass and temperature will be discussed in the future works.

\begin{figure}
\center{
\includegraphics[scale=0.9]{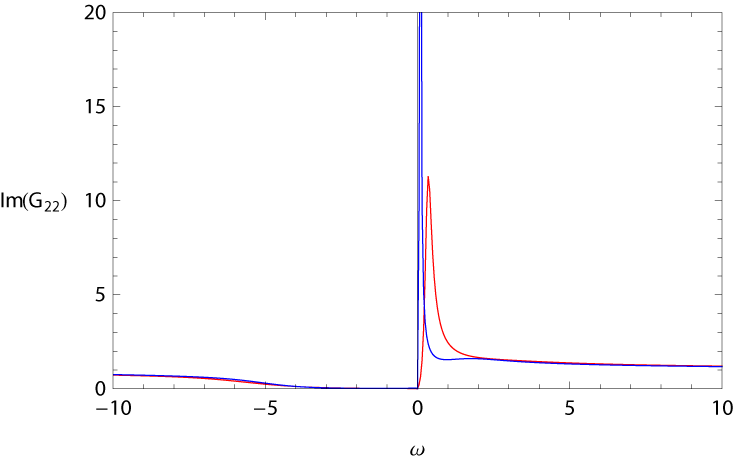}\hspace{1cm}
\includegraphics[scale=0.9]{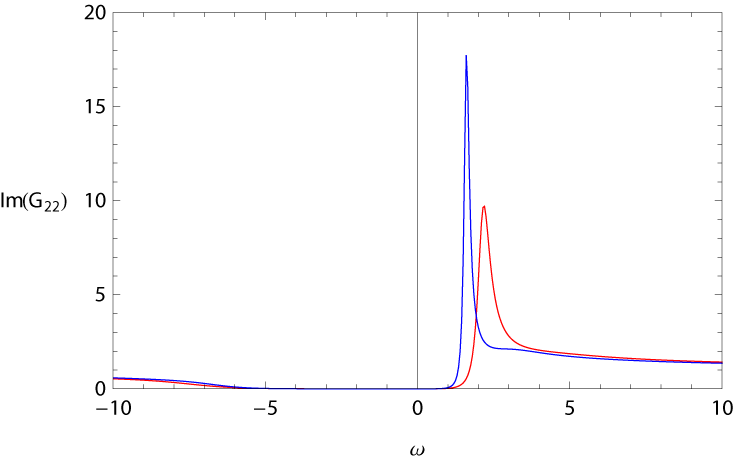}\\ \hspace{1cm}
\caption{\label{GeneralB} Spectral function $Im G_{22}$ at $k=2.2<\mu_{q}$ (left plot) and $k=4.0>\mu_{q}$ (right plot)
with different Gauss-Bonnet coupling constant $\alpha$ ($\alpha=0.09$ for red line and $\alpha=-0.19$ for blue line)
for $m_{\zeta}=0$ and $q=1$ ($\mu_{q}=\sqrt{6}$).}}
\end{figure}

Here we will do several quick checks on the consistency of our numerics.
In FIG. \ref{GeneralB}, we show the spectral function $Im G_{22}$ at $k=2.2<\mu_{q}$ (left plot) and $k=4.0>\mu_{q}$ (right plot)
for $m_{\zeta}=0$ and $q=1$ ($\mu_{q}=\sqrt{6}$).
Firstly, the divergence in the vacuum turns out to be a peak of finite size at $\omega+\mu_{q}\approx \pm k$.
In addition, independent of the parameter $\alpha$, for a fixed large $k\gg \mu_{q}$,
$Im G_{22}$ is roughly zero in the region $\omega+\mu_{q}\in (-k,k)$
and asymptote to $1$ as $|\omega|\rightarrow \infty$,
which recovers the behavior in the vacuum.
It is consistent with the Green function $\mathcal{G}$ in pure $AdS_{5}$ background (Eq. (\ref{GreenFpureAdS})).
The height and width of the peak vary with the GB coupling constant $\alpha$.
When $\alpha$ increases, the peak becomes sharper and narrower.
We also note that for $k=2.2<\mu_{q}$ (left plot),
the deviation from the vacuum behavior becomes significant.
For some more concrete investigations, we will discuss some specific properties of the spectral function in the subsequent subsection.

\subsection{Fermi surface and the dispersion relation}

\begin{figure}
\center{
\includegraphics[scale=0.9]{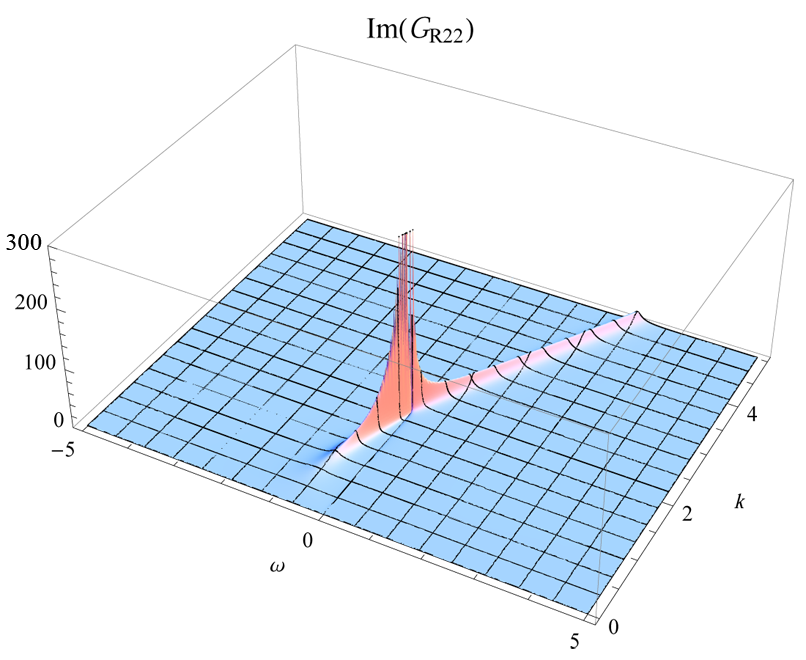}\hspace{1cm}
\caption{\label{3D} 3d plot of $Im G_{22}(\omega,k)$ for $\alpha=-0.19$,
$m_{\zeta}=0$ and $q=1$ ($\mu_{q}=\sqrt{6}$).
A sharp quasi-particle-like peak occurs near $k_{F}\approx 2.07$,
indicating a Fermi surface.
}}
\end{figure}

\begin{figure}
\center{
\includegraphics[scale=0.9]{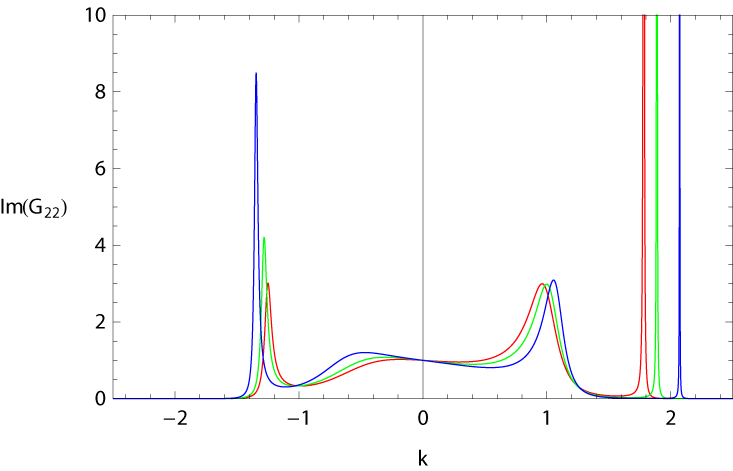}\hspace{1cm}
\caption{\label{fermiS} The plot of $Im G_{22}(k)$ for $\omega=-0.000001$
with different $\alpha$ (blue line for $\alpha=-0.19$,
green for $\alpha=0$ and red for $\alpha=0.09$). Here, $q=1$ ($\mu_{q}=\sqrt{6}$).
}}
\end{figure}

In this subsection, we will focus on the dispersion relation between small $\tilde{k}=k-k_{F}$ and $\omega$, a characteristic of Fermi or non-Fermi liquid.
In order to achieve this goal, firstly, we have to find the Fermi surface.
As is known to all, the fermion is created near the Fermi surface and so it should have a long lifetime.
Therefore, when the energy equals the Fermi energy and the momentum equals the Fermi momentum ($k_{F}$),
the spectral function of this system should have a sharp quasi-particle peak.
In this paper, we will adopt the conventions \cite{HongLiuNon-Fermi,StringQEFL},
where energy equal to the Fermi energy corresponds to the frequency vanishing.
Therefore, in order to find the Fermi surface, we can solve the equations (\ref{DiracEFCGB1}) and (\ref{DiracEFCGB2}) numerically with the boundary conditions (\ref{GatTip}).
For definiteness, we firstly focus on the cases of $q=1$ (with $\mu_{q}=\sqrt{6}$) in this subsection
and consider the cases of charge dependence in the next subsection.
For $\alpha=-0.19$, we obtain a sharp quasi-particle-like peak
near $k_{F}=2.071564$ (Fig.\ref{3D} and Fig.\ref{fermiS}).
Similarly, we can also get different fermi momentums corresponding to different $\alpha$, such as $k_{F}\approx 1.7821$ for $\alpha=0.09$, $k_{F}\approx 1.8770$ for $\alpha=0.01$, $k_{F}\approx 1.8879$ for $\alpha=0.0001$ and $k_{F}\approx 1.8880$ for $\alpha=0$ (Fig.\ref{fermiS}).
We note that the fermi momentum is $\alpha$-dependent.

\begin{figure}
\center{
\includegraphics[scale=0.9]{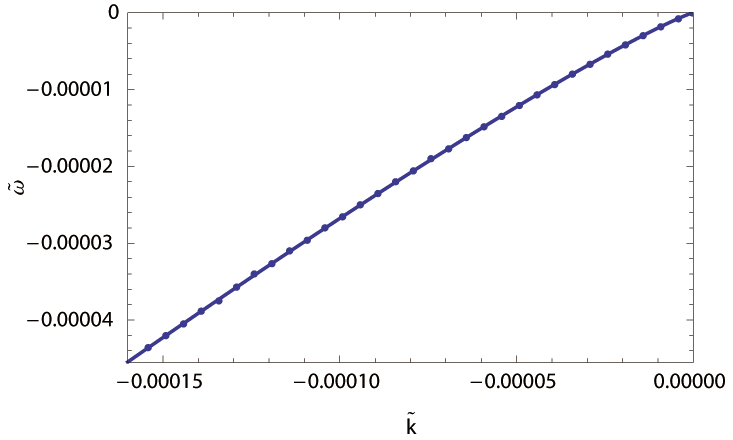}\hspace{0.1cm}
\includegraphics[scale=0.9]{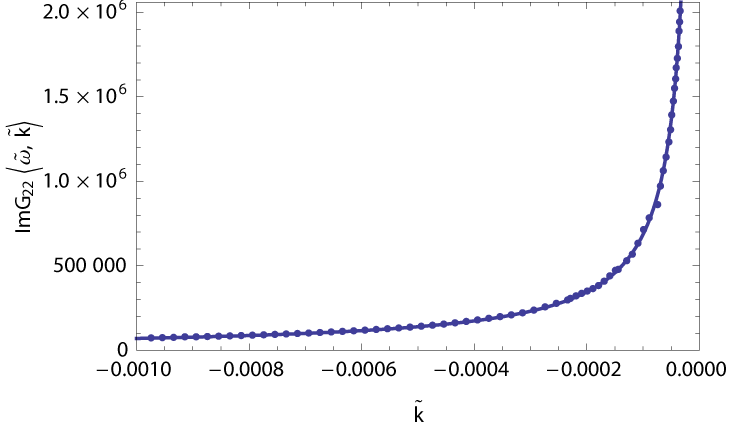}\\ \hspace{0.1cm}}
\center{
\includegraphics[scale=0.9]{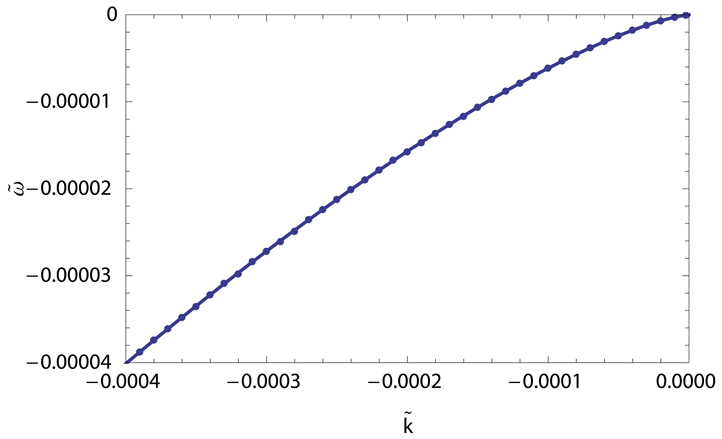}\hspace{0.1cm}
\includegraphics[scale=0.9]{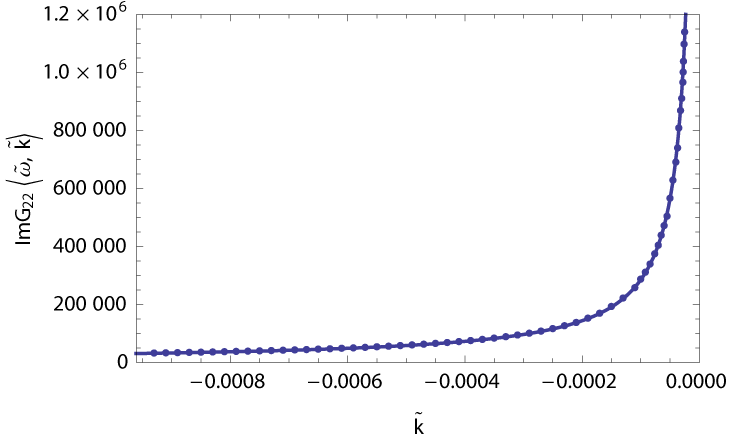}\\ \hspace{0.1cm}}
\center{
\includegraphics[scale=0.9]{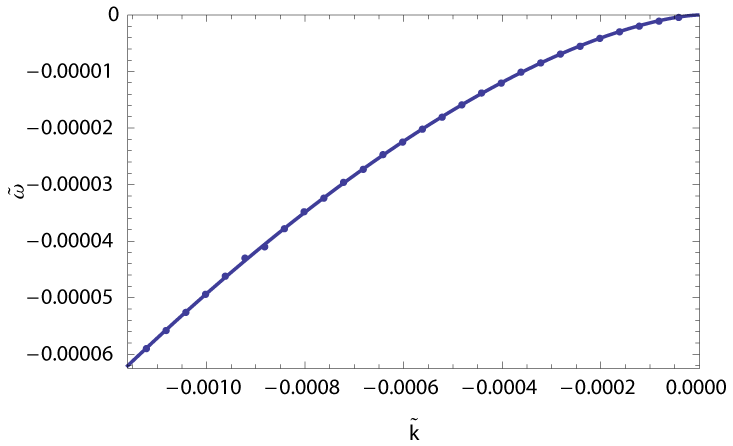}\hspace{0.1cm}
\includegraphics[scale=0.9]{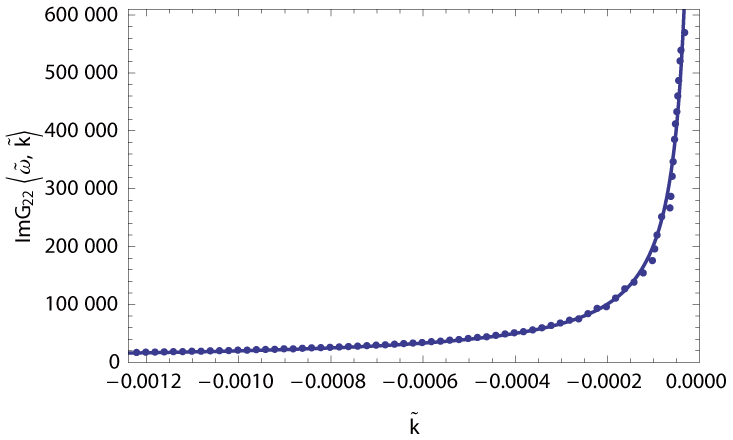}\\ \hspace{0.1cm}
\caption{\label{dispersion}Left plots: The dispersion relation between $\tilde{k}$ and $\tilde{\omega}$,
where $\tilde{k}\equiv k-k_{F}$ and $\tilde{\omega}(\tilde{k})$ is the location of the maximum of the quasi-particle-like peak. Right plots: The relation between the height of $Im G_{22}$ at the maximum and $\tilde{k}$.
From top to bottom, $\alpha=-0.19$, $0$ and $0.09$, respectively.
We fit the data as follows: $\tilde{\omega}\approx -0.855399 (-\tilde{k})^{1.12612}$
and $Im G_{22}(\tilde{\omega},\tilde{k})\approx 75.7802 (-\tilde{k})^{-0.989264}$ for $\alpha=-0.19$,
$\tilde{\omega}\approx -1.59304 (-\tilde{k})^{1.35345}$
and $Im G_{22}(\tilde{\omega},\tilde{k})\approx 31.9054 (-\tilde{k})^{-0.987734}$ for $\alpha=0$
and $\tilde{\omega}\approx -2.21699 (-\tilde{k})^{1.55103}$
and $Im G_{22}(\tilde{\omega},\tilde{k})\approx 19.2059 (-\tilde{k})^{-1.00533}$ for $\alpha=0.09$.
Here, $q=1$ ($\mu_{q}=\sqrt{6}$).}}
\end{figure}

Now, we can move on to investigate the behavior of $Im G_{22}$
in the region of small $\tilde{k}=k-k_{F}$ and $\omega$.
By fitting the data, we find that there exists a dispersion relation between $\tilde{k}$ ($\tilde{k}\rightarrow 0_{-}$) and $\tilde{\omega}(\tilde{k})$ (Fig.\ref{dispersion}), $i.e.$,
\begin{eqnarray} \label{Ldispersion}
\tilde{\omega}(\tilde{k})\sim \tilde{k}^{\delta},~~~
\end{eqnarray}
where $\delta\approx 1.13$ for $\alpha=-0.19$, $\delta\approx 1.35$ for $\alpha=0$ and $\delta\approx 1.55$ for $\alpha=0.09$.

In addition, we can also find that the scaling behavior of the height of $ImG_{22}$ at the maximum as follows:
\begin{eqnarray} \label{ScalingHeight}
Im G_{22}(\tilde{\omega},\tilde{k})\sim \tilde{k}^{-\beta},~~~\beta\approx 1,
\end{eqnarray}
for all $\alpha$ (Fig.\ref{dispersion}).

Here come some comments on the above two scaling behaviors.
They are different from the landau Fermi liquid, which has exponents $\delta=\beta=1$.
Therefore, the system can be candidates for holographic dual of generalized non-Fermi liquids.
However, we also note that when the Gauss-Bonnet coupling constant $\alpha$ is approaching the lower bound,
the parameter $\delta$ decreases,
indicating the behavior of this system is more similar to that of the landau Fermi liquid.
It seems that the GB term can also describe the dual field theories with both the weak-coupling and the strong-coupling.

In fact, when the Fermi momentum $k_{F}$ has been determined numerically,
the scaling exponents $\delta$ in the dispersion relation (\ref{Ldispersion}) can also be computed
by the analytical method developing in the Ref.\cite{HongLiuAdS2}\footnote{For details, please refer to the Ref.\cite{HongLiuAdS2}. We also give a brief summary in Appendix A.}.
We should compare the numerical result with that obtained by the analytical method.
The result summarized in Table \ref{Edelta}. From Table \ref{Edelta}, we can see that
the numerical result is agree well with the analytical that.

\begin{widetext}
\begin{table}[ht]
\begin{center}

\begin{tabular}{|c|c|c|c|c|c|c|}
         \hline
$~~\alpha~~$ &~~$-0.19$~~&~~$0$~~&~~$0.09$~~
          \\
        \hline
~~$\delta$ (numerical result)~~ & ~~$1.12612$~~ & ~~$1.35345$~~ & ~~$1.55103$~~
          \\
        \hline
~~$\delta$ (analytical result)~~ & ~~$1.14423$~~ & ~~$1.38474$~~ & ~~$1.59727$~~
          \\
        \hline
\end{tabular}
\caption{\label{Edelta} The scaling exponents $\delta$ for different GB parameter
$\alpha$ numerically and analytically.}

\end{center}
\end{table}
\end{widetext}

\subsection{Charge dependence}

As observed in Ref.\cite{HongLiuNon-Fermi,HongLiuAdS2}, the Fermi momentum $k_{F}$ increases
as we amplify the charge $q$. These features are still preserved in the charged Gauss-Bonnet black hole
for fixed GB parameter $\alpha$. We list the values of the Fermi momentum $k_{F}$
for a few other values of charge $q$ for the primary Fermi surface
in Table \ref{FermiM}\footnote{Since the peak of $G_{22}$ becomes sharper as the charge $q$ increases,
we can pin down the Fermi surface to more digit for larger $q$.}.

\begin{figure}
\center{
\includegraphics[scale=0.9]{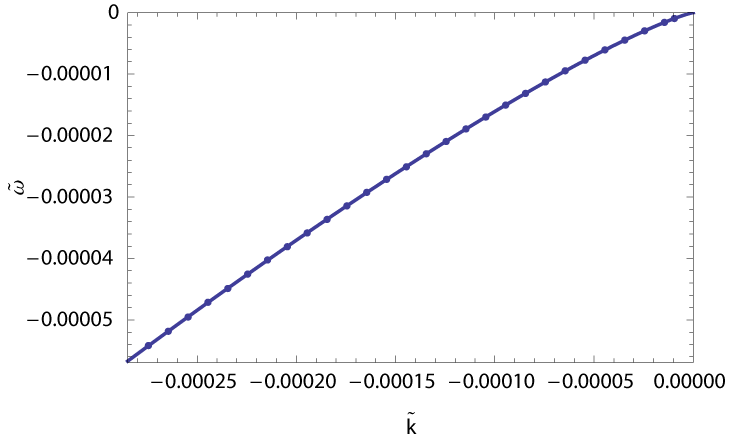}\hspace{0.1cm}
\includegraphics[scale=0.9]{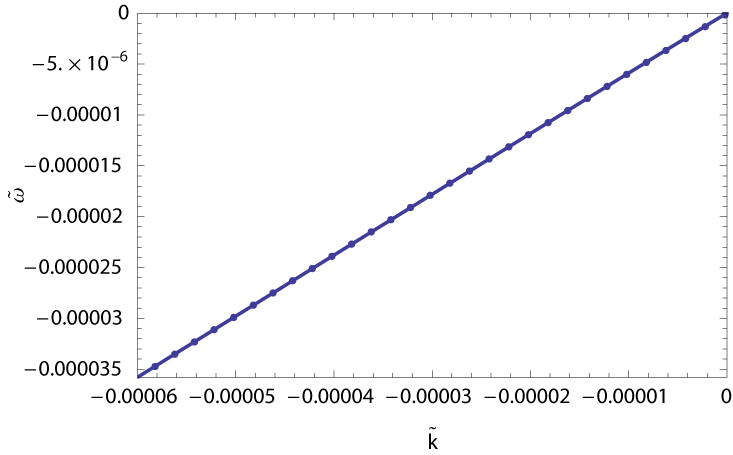}\\ \hspace{0.1cm}
\caption{\label{dispersionq12}The dispersion relation between $\tilde{k}$ and $\tilde{\omega}$ for $q=1.2$
(left plot for $\alpha=0.09$ and right plot for $\alpha=-0.19$). We fit the data as follows:
$\tilde{\omega}\approx -1.04515 (-\tilde{k})^{1.20323}$ for $\alpha=0.09$ and
$\tilde{\omega}\approx -0.642177 (-\tilde{k})^{1.0076}$ for $\alpha=-0.19$.
They show that for $q=1.2$, the scaling exponent $\delta$ decreases
with the decrease of the values of GB parameter $\alpha$
and asymptote to $1$ for smaller $\alpha$.}}
\end{figure}

After the Fermi momentum $k_{F}$ is determined numerically,
we can compute the scaling exponent $\delta$ by using Eqs. (\ref{LdispersionA}) and (\ref{nuk}).
The results are showed in Table \ref{delta}. For comparison,
we also present the numerical fitting for $q=1.2$ and $\alpha=-0.19,~0.09$
in Fig.\ref{dispersionq12}\footnote{Another scaling exponent $\beta$ is still approximate to $1$ independent of $\alpha$ and $q$.}.
From Table \ref{delta} and Fig.\ref{dispersionq12}, we can see that for fixed GB parameter $\alpha$,
with the increase of charge $q$, the scaling exponent $\delta$ decreases rapidly
and will asymptote to $1$ for larger values of $q$.
It is in agreement with that found in Ref.\cite{HongLiuNon-Fermi}.
When the charge $q$ is increased to certain values (for example, $q=1.2$),
we find that the scaling exponent $\delta$ decreases with the decrease of
the values of GB parameter $\alpha$ and asymptote to $1$ for smaller $\alpha$.
While for larger charge $q$ (for example, $q=1.5$),
the scaling exponent $\delta\approx 1$ independent of the $\alpha$.
For furthermore exploration on how the charge $q$ and GB parameter $\alpha$
affect together the scaling exponent $\delta$, we leave it for future work.

\begin{widetext}
\begin{table}[ht]
\begin{center}

\begin{tabular}{|c|c|c|c|c|c|c|}
        \hline
$~~ ~~$ &~~$q=0.5$~~&~~$q=1.2$~~&~~$q=1.5$~~
          \\
        \hline
$~~\alpha=-0.19~~$ & ~~$0.87$~~ &~~$2.581414$~~&~~$3.3570706$~~
          \\
        \hline
$~~\alpha=0~~$ & ~~$0.81$~~ & ~~$2.349913$~~ & ~~$3.0572678$~~
          \\
        \hline
$~~\alpha=0.09~~$ & ~~$0.77$~~ & ~~$2.212214$~~ & ~~$2.8735578$~~
          \\
        \hline
\end{tabular}
\caption{\label{FermiM} The Fermi momentum $k_{F}$ for different charge $q$ and GB parameter $\alpha$.}

\end{center}
\end{table}
\end{widetext}

\begin{widetext}
\begin{table}[ht]
\begin{center}

\begin{tabular}{|c|c|c|c|c|c|c|}
        \hline
$~~ ~~$ &~~$q=0.5$~~&~~$q=1.2$~~&~~$q=1.5$~~
          \\
        \hline
$~~\alpha=-0.19~~$ & ~~$3.41726$~~ &~~$1$~~&~~$1$~~
          \\
        \hline
$~~\alpha=0~~$ & ~~$4.38389$~~ & ~~$1.06558$~~ & ~~$1$~~
          \\
        \hline
$~~\alpha=0.09~~$ & ~~$5.68267$~~ & ~~$1.22051$~~ & ~~$1$~~
          \\
        \hline
\end{tabular}
\caption{\label{delta} The scaling exponent $\delta$ for different charge $q$
and GB parameter $\alpha$ (analytical result).}

\end{center}
\end{table}
\end{widetext}

\subsection{Particle-Hole (A)symmetry}

\begin{figure}
\center{
\includegraphics[scale=0.6]{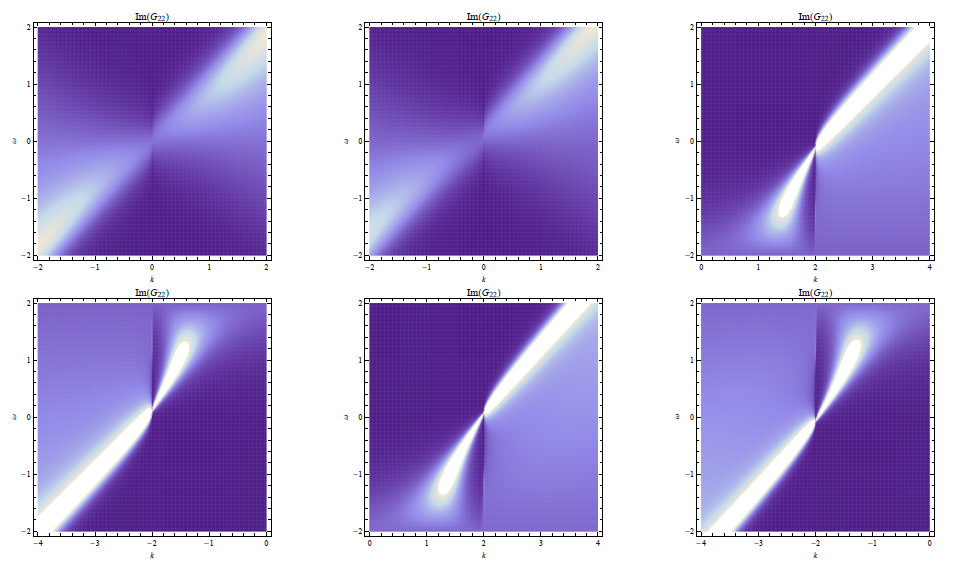}\hspace{1cm}
\caption{\label{ParticleHole} Density plots of the spectral function for $T=0$.
Lighter regions have higher value of the function.
The three panels above correspond to $\alpha=-0.19$ and $q=0$, $\alpha=-0.19$ and $q=0.1$, $\alpha=0$ and $q=1$ (from left to right).
The three panels below correspond to $\alpha=0$ and $q=-1$, $\alpha=-0.19$ and $q=1$, $\alpha=-0.19$ and $q=-1$ (from left to right).}}
\end{figure}

As mentioned in subsection A, when the background geometry is pure $AdS_{5}$ and $q=0$,
the spectral function has a Particle-hole symmetry, but when $q\neq 0$, the Particle-hole asymmetry presents.
The similar case does occur in the background of charged Gauss-Bonnet black hole (Fig.\ref{ParticleHole}).
In Fig.\ref{ParticleHole}, the last two panels below show that
the spectral function behavior is not symmetrical about the Fermi point (Particle-hole asymmetry) for $\alpha=-0.19$ and $q=1$ or $q=-1$.
Moreover, we also find that when $q$ is restored to $0$, the system regains the Particle-hole symmetry
and the asymmetry slowly becomes obvious with $q$ increasing (the first and second panel above in Fig.\ref{ParticleHole}).
Thus, we can reasonably conclude that
whether the system has the Particle-hole symmetry or asymmetry is related to the different values of $q$.
Some similar conclusions have also been pointed out in the investigations on fermions in charged BTZ black hole \cite{FermionsBTZBH}.
In addition, we also notice that the Gauss-Bonnet coupling constant $\alpha$ has less effects on the (a)symmetry of the system.
For comparison, we present the plot for $\alpha=0$ and $q=1$ or $q=-1$ (the last panel above and the first panel below in Fig.\ref{ParticleHole}).

\section{Conclusions and discussion}

We have studied the main features of the fermions in charged Gauss-Bonnet black hole
for zero temperature limit and massless fermions by AdS/CFT correspondence.
The general behavior of the spectral function is similar with the case of RN black hole \cite{HongLiuNon-Fermi}.
However, the Gauss-Bonnet coupling constant $\alpha$ changes the shape of the spectral function.
Especially, near the quasi-particle like peak, the effect the $\alpha$ exerts on the shape of the spectral function is more significant.
Therefore, it is interesting to further understand the behavior of Green¡¯s functions in the
Quasi-topological gravity or the case with Weyl corrections,
which has more physical contents.

Furthermore, we especially focus on the dispersion and the Particle-hole asymmetry.
Their properties can be summarized as follows.
Generally, at $q=1$, we find a dispersion relation $\tilde{\omega}(\tilde{k})\sim \tilde{k}^{\delta}$
and the scaling exponent $\delta\neq 1$, indicating this system is non-Fermi liquid.
Therefore, the system can be candidates for holographic dual of generalized non-Fermi liquids.
More importantly, the behavior of this system increasingly similar to that of the Landau Fermi liquid
when $\alpha$ is approaching its lower bound.
It seems that the GB term can also describe the dual field theories with both
the weak-coupling and the strong-coupling.
Also we discuss briefly the cases of the charge dependence.
At larger values of $q$, new scaling behavior appears. For instance,
for certain values of charge $q$ (for example, $q=1.2$),
the scaling exponent $\delta$ decreases with the decrease of
the values of GB parameter $\alpha$ and asymptote to $1$ for smaller $\alpha$.
While for larger charge $q$ (for example, $q=1.5$),
the scaling exponent $\delta\approx 1$ independent of the $\alpha$.
Another important characteristic of this system is the Particle-hole asymmetry.
From the density plot of the spectral function,
we find that for zero temperature limit,
the Particle-hole (a)symmetry is controlled by the parameter $q$
and the Gauss-Bonnet coupling constant $\alpha$ has less effects on it.

In addition, other features, such as the log period, studies in Ref. \cite{HongLiuNon-Fermi}, are still preserved,
which are determined by the geometry $AdS_{2}$ near the horizon.
In the future works, we will address the particulars of such properties.
In order to searching for the effect of the higher curvature corrections on the spectral function of the fermions,
we can study the fermions in Quasi-topological gravity or the case with Weyl corrections in parallel.

%Especially, we should point out that in Refs. \cite{Weyl2,WeylJPWu},
%an interesting phenomenon has been revealed that the dual theory can vary from weak-coupling to strong-coupling as the Weyl coupling parameter.

\begin{acknowledgments}

I would like to thank my advisor, Prof. Yongge Ma, for his encouragement.
And I am also grateful to Prof. Yi Ling, Dr. Hongbao Zhang, Dr. Wei-Jie Li,
Dr. Xiangdong Zhang, Dr. Huaisong Zhao and Yue Cao for their useful discussions.
In addition, I am also thank J. Shock for his pointing out some type mistakes and his comments.
This work is partly supported by NSFC(No.10975017) and the Fundamental Research
Funds for the central Universities.

\end{acknowledgments}

\begin{appendix}

\section{Analytical treatment}

In this Appendix, we will give a brief summary on the analytic treatment
in the low frequency limit developed in Ref.\cite{HongLiuAdS2}.
As revealed in Ref.\cite{HongLiuAdS2}, the dispersion relation can be given
by\footnote{The equation (93) in Ref.\cite{HongLiuAdS2}.}
\begin{eqnarray} \label{LdispersionA}
\tilde{\omega}(\tilde{k})\propto \tilde{k}^{\delta}, \quad {\rm with} \quad \delta = \begin{cases} \frac{1}{2 \nu_{k_F}} & \nu_{k_F} < \frac{1}{2}\cr
            1 & \nu_{k_F} > \frac{1}{2}
            \end{cases}.
\end{eqnarray}

While $\nu_{k_{F}}$ can be calculated by the equation (57) in Ref.\cite{HongLiuAdS2}
\begin{eqnarray} \label{nuk}
\nu_{k}=\frac{g_{F}q}{\sqrt{2d(d-1)}}\sqrt{\frac{2m_{\zeta}^{2}}{g_{F}^{2}q^{2}}
+\frac{d(d-1)}{(d-2)^{2}}\frac{k^{2}}{\mu_{q}^{2}}-1}~~.
\end{eqnarray}

In our conventions, $g_{F}=2$. The above expressions was derived by the fact
that the geometry near the horizon is $AdS_{2}\times \mathbb{R}^{d-1}$.
As discussed in Section\ref{CBHinGBG}, the geometry near the horizon is also
$AdS_{2}\times \mathbb{R}^{3}$ in Gauss-Bonnet gravity, independent of $\alpha$.
Therefore, in order to obtain the dispersion relation,
we only need to work out numerically the Fermi momentum $k_{F}$,
which is controlled by UV physics.

However, we can also obtain the range of $k_{F}$ by WKB
analysis\footnote{This analysis only applies to the Reissner-Nordstr$\ddot{o}$m
AdS black hole but not Gauss-Bonnet AdS black hole.} \cite{HongLiuAdS2}.
As observed in Ref.\cite{HongLiuAdS2}, when $m_{\zeta}=0$, the range for allowed $k_{F}$
is\footnote{The equation (110) in Ref.\cite{HongLiuAdS2}.}
\begin{eqnarray} \label{RangekF}
\frac{d-2}{\sqrt{d(d-1)}}\leq \frac{k_{F}}{\mu_{q}}\leq 1.
\end{eqnarray}

In order to test the robustness of our numerical result on the Fermi momentum.
We list the values of the Fermi momentum $k_{F}$ determined numerically and the range of $k_{F}$
determined by Eq.(\ref{RangekF}) in Table \ref{kF} for $\alpha=0$ and different $q$.
From Table \ref{kF}, we can see that our numerical result
is well within the interval determined by Eq.(\ref{RangekF}).

\begin{widetext}
\begin{table}[ht]
\begin{center}

\begin{tabular}{|c|c|c|c|c|c|c|}
        \hline
$~~q~~$ &~~$0.5$~~&~~$1$~~&~~$1.2$~~&~~$1.5$~~
          \\
        \hline
$~~k_{F}$(numerical)~~ & ~~$0.81$~~ &~~$1.8880$~~&~~$2.349913$~~&~~$3.0572678$~~
          \\
        \hline
$~~k_{F}$(WKB)~~ & ~~$[0.707,1.225]$~~ &~~$[1.414,2.450]$~~&~~$[1.697,2.939]$~~&~~$[2.121,3.674]$~~
          \\
        \hline
\end{tabular}
\caption{\label{kF} The Fermi momentum $k_{F}$ determined numerically
and the range of $k_{F}$ obtained by WKB analysis.}

\end{center}
\end{table}
\end{widetext}

\end{appendix}


\begin{thebibliography}{99}

%%%%%%%%%%%%%%%%% The AdS/CFT correspondence

\bibitem{Maldacena1997}
J. M. Maldacena, The large N limit of superconformal field theories and supergravity,
Adv. Theor. Math. Phys. 2, 231 (1998) [Int. J. Theor. Phys. 38, 1113 (1999)],
[hep-th/9711200].

\bibitem{Gubser1998}
S. S. Gubser, I. R. Klebanov, and A. M. Polyakov, Gauge theory correlators from non-critical
string theory, Phys. Lett. B \textbf{428} (1998) 105¨C114, [hep-th/9802109].

\bibitem{Witten1998}
E. Witten, Anti-de Sitter space and holography, Adv. Theor. Math. Phys. 2 (1998) 253¨C291,
[hep-th/9802150].



%%%%%%%%%%%%%%%%% HongLiu Non-Fermi Liquids

\bibitem{HongLiuNon-Fermi}
H. Liu, J. McGreevy and D. Vegh, Non-Fermi Liquids
from Holography, [arXiv:0903.2477].



%%%%%%%%%%%%%%%%% String Theory, Quantum Phase Transitions and the Emergent Fermi-Liquid

\bibitem{StringQEFL}
M. $\check{C}$ubrovi$\acute{c}$, J. Zaanen, K. Schalm, String Theory, Quantum Phase Transitions and the Emergent Fermi-Liquid,
Science \textbf{325}, 439 (2009), [arXiv:0904.1993].

%%%%%%%%%%%%%%%%%% Fermions in BTZ black hole

\bibitem{FermionsBTZBH}
D. Maity, S. Sarkar, B. Sathiapalan, R.Shankar and N. Sircar, Properties of CFTs dual to Charged BTZ black-hole, Nucl. Phys. B \textbf{839}, 526 (2010), [arXiv:0909.4051].


%%%%%%%%%%%%%%%%%% Other related holographic fermions models

\bibitem{HNFMagneticFieldBasu}
P. Basu, J. Y. He, A. Mukherjee, H. H. Shieh, Holographic Non-Fermi Liquid in a Background Magnetic Field, Phys. Rev. D \textbf{82}, 044036 (2010), [arXiv:0908.1436].

\bibitem{StrangeMetallicHartnoll}
S. A. Hartnoll, J. Polchinski, E. Silverstein, D. Tong, Towards strange metallic holography, JHEP \textbf{1004}, 120 (2010), [arXiv:0912.1061].

\bibitem{SemiHFLPolchinski}
T. Faulkner, J. Polchinski, Semi-Holographic Fermi Liquids, [arXiv:1001.5049].

\bibitem{HNFLMagneticFGubankova}
E. Gubankova, J. Brill, M. Cubrovic, K. Schalm, P. Schijven, J. Zaanen, Holographic fermions in external magnetic fields, [arXiv:1011.4051].

\bibitem{HFLDynamicalGap}
M. Edalati, R. G. Leigh, K. W. Lo, P. W. Phillips, Dynamical Gap and Cuprate-like Physics from Holography, [arXiv:1012.3751].

\bibitem{HFLDipoleCoupling}
D. Guarrera, J. McGreevy, Holographic Fermi surfaces and bulk dipole couplings, [arXiv:1102.3908].


%%%%%%%%%%%%%%% GB holographic superconductors

\bibitem{GBHS1}
R. Gregory, S. Kanno and J. Soda, Holographic Superconductors with Higher Curvature Corrections, JHEP \textbf{0910}, 010 (2009), [arXiv:0907.3203].

\bibitem{GBHS2}
Q. Pan, B. Wang, E. Papantonopoulos, J. Oliveira and A. B. Pavan, Holographic Superconductors with various condensates in Einstein-Gauss-Bonnet gravity,
Phys. Rev. D \textbf{81}, 106007 (2010), [arXiv:0912.2475].

\bibitem{GBHS3}
Q. Pan and B. Wang, General holographic superconductor models with Gauss-Bonnet corrections, Phys. Lett. B \textbf{693}, 159 (2010), [arXiv:1005.4743].

\bibitem{GBHS4}
R. G. Cai, Z. Y. Nie and H. Q. Zhang, Holographic p-wave superconductors from Gauss-Bonnet gravity, Phys. Rev. D \textbf{82}, 066007 (2010) [arXiv:1007.3321].

\bibitem{GBHS5}
L. Barclay, R. Gregory, S. Kanno and P. Sutcliffe, Gauss-Bonnet Holographic Superconductors, JHEP \textbf{1012}, 029 (2010), [arXiv:1009.1991].


\bibitem{GBHS6}
X. H. Ge, B. Wang, S. F. Wu, G. H. Yang, Analytical study on holographic superconductors in external magnetic field, JHEP \textbf{1008}, 108 (2010), [arXiv:1002.4901].

\bibitem{GBHS7}
Y. Brihaye, B. Hartmann, Holographic Superconductors in 3+1 dimensions away from the probe limit, Phys. Rev. D \textbf{81}, (2010) 126008, [arXiv:1003.5130].


%%%%%%%%%%%%%%% Black hole in Quasi-topological Gravity

\bibitem{QT}
J. Oliva and S. Ray, A new cubic theory of gravity in five dimensions: Black hole, Birkhoff's theorem and C-function, Class. Quant. Grav. \textbf{27}, 225002 (2010), [arXiv:1003.4773].


\bibitem{QTblackhole}
R. C. Myers and B. Robinson, Black Holes in Quasi-topological Gravity, JHEP \textbf{1008}, 067 (2010), [arXiv:1003.5357].


%%%%%%%%%%%%%%% Holographic superconductors in Qusi-topological Gravity

\bibitem{HSinQT1}
X. M. Kuang, W. J. Li, Y. Ling, Holographic Superconductors in Quasi-topological Gravity, JHEP \textbf{1012}, 069 (2010), [arXiv:1008.4066].

\bibitem{HSinQT2}
M. Siani, Holographic Superconductors and Higher Curvature Corrections, JHEP \textbf{1012}, 035 (2010), [arXiv:1010.0700].


%%%%%%%%%%%%%%%%%% Weyl corrections

\bibitem{Weyl1}
A. Ritz and J. Ward, Weyl corrections to holographic conductivity, Phys. Rev. D \textbf{79},
066003 (2009), [arXiv:0811.4195].

\bibitem{Weyl2}
R. C. Myers, S. Sachdev and A. Singh, Holographic quantum critical transport without selfduality,
[arXiv:1010.0443].


%%%%%%%%%%%%%%%%% Weyl Holographic superconductors

\bibitem{WeylHS}
J. P. Wu, Y. Cao, X. M. Kuang, W. J. Li, The 3+1 holographic superconductor with Weyl corrections, Phys. Lett. B \textbf{697}, 153 (2011), [arXiv:1010.1929].





%%%%%%%%%%%%%%% GB black hole

\bibitem{GBblackhole1}
D. G. Boulware and S. Deser, String-Generated Gravity Models, Phys. Rev. Lett. \textbf{55}, 2656 (1985).

\bibitem{GBblackhole2}
R. G. Cai, Gauss-Bonnet Black Holes in AdS Spaces, Phys. Rev. D \textbf{65}, 084014 (2002), [arXiv:hep-th/0109133].


%%%%%%%%%%%%%%%% GB coupling Constraints

\bibitem{GBcouplingConstraint1}
M. Brigante, H. Liu, R. C. Myers, S. Shenker and S. Yaida, Viscosity Bound Violation in Higher Derivative Gravity, Phys. Rev. D \textbf{77}, 126006 (2008)
[arXiv:0712.0805].

\bibitem{GBcouplingConstraint2}
X. H. Ge, Y. Matsuo, F. W. Shu, S. J. Sin, T. Tsukioka, Viscosity Bound, Causality Violation and Instability with Stringy Correction and Charge,
JHEP \textbf{0810}, 009 (2008), [arXiv:0808.2354].

\bibitem{GBcouplingConstraint3}
M. Brigante, H. Liu, R. C. Myers, S. Shenker and S. Yaida, Viscosity Bound and Causality Violation, Phys. Rev. Lett. \textbf{100}, 191601
(2008) [arXiv:0802.3318].

\bibitem{GBcouplingConstraint4}
A. Buchel and R. C. Myers, Causality of Holographic Hydrodynamics, JHEP \textbf{0908}, 016 (2009) [arXiv:0906.2922].

\bibitem{GBcouplingConstraint5}
D. M. Hofman, Higher Derivative Gravity, Causality and Positivity of Energy in a UV complete QFT, Nucl. Phys. B \textbf{823}, 174 (2009) [arXiv:0907.1625].

\bibitem{GBcouplingConstraint6}
J. de Boer, M. Kulaxizi and A. Parnachev, $AdS_{7}/CFT_{6}$, Gauss-Bonnet Gravity, and Viscosity Bound, JHEP \textbf{1003}, 087 (2010) [arXiv:0910.5347].

\bibitem{GBcouplingConstraint7}
X. O. Camanho and J. D. Edelstein, Causality constraints in AdS/CFT from conformal collider physics and Gauss-Bonnet gravity, JHEP \textbf{1004}, 007 (2010) [arXiv:0911.3160].

\bibitem{GBcouplingConstraint8}
A. Buchel, J. Escobedo, R. C. Myers, M. F. Paulos, A. Sinha and M. Smolkin, Holographic GB gravity in arbitrary dimensions, JHEP \textbf{1003},
111 (2010) [arXiv:0911.4257].

\bibitem{GBcouplingConstraint9}
X. H. Ge, S. J. Sin, Shear viscosity, instability and the upper bound of the Gauss-Bonnet coupling constant,
JHEP \textbf{0905}, 051 (2009), [arXiv:0903.2527].

\bibitem{GBcouplingConstraint10}
X. H. Ge, S. J. Sin, S. F. Wu, G. H. Yang, Shear viscosity and instability from third order Lovelock gravity,
Phys. Rev. D \textbf{80}, 104019 (2009), [arXiv:0905.2675].






%%%%%%%%%%%%%%%%% Conventions

\bibitem{Conventions1}
R. M. Wald, ¡°General Relativity,¡± The University of Chicago Press.


%%%%%%%%%%%%%%%%% Photoemission

\bibitem{Photoemission}
T. Faulkner, G. T. Horowitz, J. McGreevy, M. M. Roberts, D. Vegh, Photoemission "experiments" on holographic superconductors, JHEP \textbf{1003}, 121 (2010), [arXiv:0911.3402].





%%%%%%%%%%%%%%%%% HongLiu AdS2

\bibitem{HongLiuAdS2}
T. Faulkner, H. Liu, J. McGreevy and D. Vegh, Emergent
Quantum Criticality, Fermi Surfaces, and $AdS_{2}$,
[arXiv:0907.2694].


%%%%%%%%%%%%%%%%% HongLiu Spinor

\bibitem{HongLiuSpinor}
N. Iqbal and H. Liu, Real-time response in AdS/CFT with application to spinors,
Fortsch. Phys. \textbf{57}, 367 (2009), [arXiv:0903.2596].



%%%%%%%%%%%%%%%%% HongLiu Spinor

\bibitem{HongLiuUniversality}
N. Iqbal and H. Liu, Universality of the Hydrodynamic
Limit in AdS/CFT and the Membrane Paradigm, Phys.
Rev. D \textbf{79}, 025023 (2009), [arXiv:0809.3808].



%%%%%%%%%%%%%%%%% AdS spinnor vacuum

\bibitem{GreenFpureAdS1}
M. Henningson and K. Sfetsos, Spinors and the AdS/CFT correspondence, Phys. Lett. B \textbf{431}, 63
(1998), [arXiv:hep-th/9803251].

\bibitem{GreenFpureAdS2}
W. Mueck and K. S. Viswanathan, Conformal Field Theory Correlators from Classical Field Theory on Anti-de Sitter Space II. Vector and Spinor Fields, Phys. Rev. D \textbf{58}, 106006 (1998), [arXiv:hep-th/9805145].









\end{thebibliography}
\end{document}